# Seismic Wave Scattering and Dissipation in Fractured Shales

**Hao Zhou[1], Xiaoping Jia[2*], Li-Yun Fu[3†], and Arnaud Tourin[2]**

[1] *College of Geophysics, Chengdu University of Technology, Chengdu 610059, China.*

[2] *Institut Langevin, ESPCI Paris, Université PSL, CNRS, Paris 75005, France.*

[3] *School of Geosciences, China University of Petroleum, Qingdao 266580, China.*

**Abstract**

Seismic attenuation in granular porous media is of paramount importance in rock physics and seismology. Unlike sandstones, shales are mixtures of sand grains and clays with extremely low porosity and permeability. Swelling of clays upon wetting induce micro-cracks at grain-clay interfaces and results in the strong elastic wave scattering. Such scattering prevents adequate measurements of the absorption from ballistic wave attenuations. Here we infer this intrinsic attenuation from multiply scattered waves as in seismology and ultrasonics. We find that increasing confining pressure reduces the scattering attenuation by micro-crack closure but increases surprisingly the absorption, likely due to the viscous dissipation involved with more liquids adsorbed in clays and at grain surfaces. Also, we observe that cyclic heating and cooling causes the shrinkage of clays and the growth of microcracks as well as the nucleation of macro-fractures. This leads to a predominant chaotic reverberation in this fractured shale. Numerical simulations based on X-ray tomography of the fractured sample confirm the multiple scattering behavior and reveal the increase of a characteristic length from an initial intact to a finally fractured shale. This study helps to improve acoustic techniques for multiscale exploration of gas and oil in shales and other fractured rocks.

## 1. Introduction

Determinations of seismic velocity and attenuation in granular porous rocks are of paramount importance to exploration geophysics (Bourbié et al., 1987). Attenuation is particularly sensitive to the rock property change in the presence of pore fluid (Winkler and Nur, 1982). In the laboratory experiment, we may investigate ultrasound propagation in such heterogeneous media to study different mechanisms of attenuation like *anelastic* dissipation (absorption) and *elastic* scattering, for the range of 50 kHz - 5 MHz. In nearly nonporous and impermeable granites, the attenuation is dominated by scattering due to the acoustic velocity fluctuations on the scale of grains (Fukushima et al, 2003; Zhou et al., 2021) similar to those observed in polycrystals (Weaver, 1990) or unconsolidated dry granular materials due to the heterogeneous contact network (Jia, 2004). However, in wet porous sandstones (Berea) and granular materials, partially or totally saturated by liquid, viscoelastic dissipation due to wave-induced pore fluid flow becomes the dominant mechanism of attenuation (Jones, 1986; Müller et al., 2010).

---

* Xiaoping.jia@espci.fr
† lfu@upc.edu.cn

Such a fluid flow may exhibit different behavior as a function of the distribution of pore space and the wavelength, giving rise to the macroscopic Biot's flow (Biot, 1956a, b, 1962), microscopic squirt flow (Mavko and Nur, 1975; O'Connell and Budiansky, 1977), and mesoscopic flow (Pride et al., 2004), respectively. When the characteristic void size in pore spaces is larger than ~ 10 µm, the pore fluid can flow freely (Kowalski, 2003). Otherwise, the moisture in the pore space, usually a mixture of gas and water or water solution, is likely to bond to the solid particles by various forces (chemical, physical-chemical, physical-mechanical bounds). In such cases, absorption mechanisms associated with small amounts of liquids closely bound to the host particle surfaces (Tittmann et al., 1980) or trapped by grain asperities (Brunet et al., 2008) should become predominant. However, this issue is still not well understood in exploration geophysics (Jones, 1986).

Attenuation measurements in shales are more complicated than in ordinary rock materials since scattering and intrinsic attenuations seem to be equally important, and the radii of voids are usually on the nanometer scale. Shale is a common sedimentary rock that acts as a source and seal in conventional oil and gas reservoirs (Johnston, 1987), and has become well-known in recent years for shale gas (Wang et al., 2014). It consists of very fine-grained minerals, such as quartz, calcite, pyrite, with an average particle size of up to a few microns, and most importantly, clay minerals (Grim, 1953) (see Fig. 1a2). These clay minerals form layers due to the compaction of the overburden during the diagenesis process. As such, shale is characterized by its tendency to split into thin layers (laminae) less than one centimeter in thickness, called *fissility* (Ingram, 1953). In terms of elastic properties, shales show strong anisotropy due to these layers, which can be affected by microcracks, organic matters, and the pressure-temperature conditions, leading to fractures developed along the bedding (e.g., Jones and Wang, 1981; Vernik and Nur, 1992; Vernik, 1993; Allan et al., 2016; Sayers, 2016). For ultrasonic experiments, scattering from these layers and cracks has been indirectly noted based on the coherent wave measurement (Zhubayev et al., 2016); however, a quantitative assessment of scattering is still lacking. For dissipation measurement, wave-induced flows from microscopic to macroscopic do not seem to be relevant in shales, because fluids confined in the nanoscale intergranular pore space are subjected to strong capillary forces and other surface effects. Subsequently, they cannot flow freely (Ortega et al., 2009).

As mentioned above, efficient inverse methods are generally not available in exploration geophysics to separate scattering and intrinsic attenuations (Pride et al, 2004). The coherent-wave based techniques are affected by wave scattering (Toksöz et al., 1979, Quan and Harris, 1997). In seismology and ultrasonics, multiply scattered waves (coda) are not only applied to measure the wave velocity change (Snieder et al, 2002), but also employed to infer the intrinsic attenuations (absorption) via the diffusion or radiative transfer equation (e.g., Fehler et al., 1992; Lacombe et al., 2003; Page et al., 1995; Jia, 2004; Brunet et al, 2008). In practice, the analytical solutions are often not available because of the source-detector size effect and the boundary conditions in finite-size samples. Numerical simulations are then performed for the inverse purpose like Monte-Carlo methods (Zhou et al., 2021). However, such kinds of simulations may miss the underlying physics (microstructure of materials) responsible for scattering and absorption (e.g., Ghoshal and Turner, 2009; Ryzy et al., 2018).

In this work, we investigate the ultrasound scattering and absorption in room-dry and wet shales under hydrostatic loading and cyclic heating-cooling, respectively (section 2).

Finite-difference (FD) simulations of ultrasound scattering are performed on the basis of high-resolution X-ray computed tomography images of fractured shales (section 3). The absorption (intrinsic attenuation) in such macro-fractured media is inferred from a diffusion model as in a reverberant acoustic room. We show that the absorption is controlled by the viscous dissipation of small amount of trapped water in intergranular clay minerals or cracks and at grain surfaces (section 4).

## 2. Experiments

We collected several shale samples for ultrasonic experiments from the Long-ma-xi formation at a depth about 2200 m in Sichuan Province, China. A cylindrical sample is shown in Fig. 1a1 with a diameter of $D \approx 25.4$ mm and a length of $L \approx 66.5$ mm. X-ray diffraction (XRD) analysis shows that it contains about 60% quartz, 20% clay minerals, and other minor minerals (feldspar, pyrite, etc.); Fig. 1$a$2 exhibits some of these minerals. Clay minerals consist of about 80% illite, 5% smectite, and 15% chlorite. This sample is compact with a porosity of about 5% and an absolute permeability of about 0.1mD. The organic matter content (TOC) is about 5% with a relatively high maturity, and the sample appears black. The bedding is parallel to the central axis of the cylindrical sample. Scanning electron microscopy (SEM) images (Fig. 1a2) illustrate a composition of fine quartz ($d \approx 5\mu m$) encapsulated and cemented by intergranular mixture of clay minerals and organic matters. The microstructure of this initial shale is generally intact (no *visible cracks*, hereinafter denote as *fractures*) with a few microcracks developed along grain boundaries.

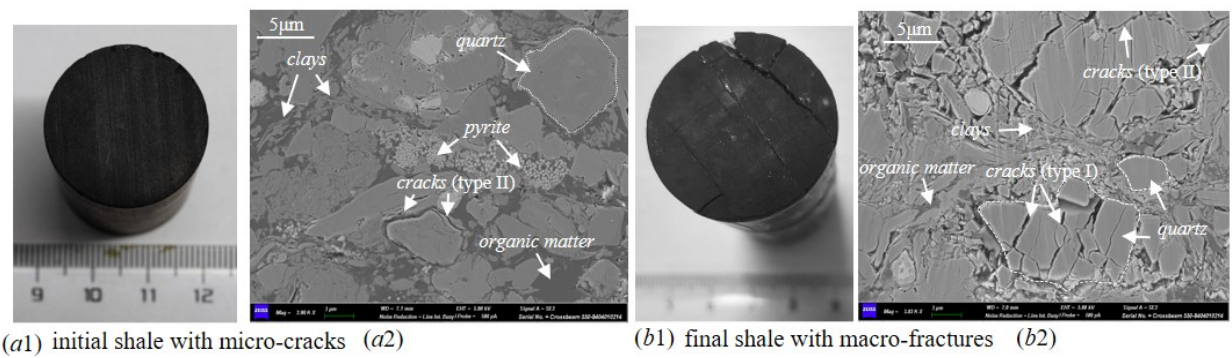

($a$1) initial shale with micro-cracks ($a$2)   ($b$1) final shale with macro-fractures ($b$2)

*Figure 1. Shale samples. (a1) The initial intact shale sample with micro-cracks with (a2) its SEM photograph). (b1) The final fractured shale sample after cyclic heating and cooling with (b2) its SEM photograph.*

### 2.1 Hydrostatic loading in room-dry and wet shales

We first performed hydrostatic loading experiments in the intact shale sample, combined with ultrasonic measurements. As shown in Fig. 2a, it is a triaxial loading cell where the confining pressure ($P_c$) is generated by an incremental injection of silicon oil into a sealed pressure chamber. The ultrasonic source and receiver transducers are encapsulated in two metal housings in which a fluid inlet is embedded to generate pore pressure. The sample and the two metal housings were coupled and bonded together, jacketed by a rubber sleeve, and placed in the pressure chamber. In this experiment, the axial pressure is equal to the confining pressure $P_c$, leading to a hydrostatic loading. During measurements, $P_c$ increases from 5 MPa to 55 MPa, without control of pore pressure. We spent one hour for stabilizing each given pressure.

As to the ultrasonic measurement, a 5-cycle sine burst source signal centered at 0.5 MHz was sent to a shear source transducer of 12 mm in diameter at the top and the transmitted ultrasonic signals were detected by the same transducer at the bottom of the sample. We have investigated the same sample in *room-dry* and *wet* conditions, respectively. *Room-dry* condition involves the sample first heated at 105°C for 24 hours and then cooled in air at room temperature (~ 20 °C) prior to testing. *Wet* condition consists of soaking the sample in water for 24 hours in order to absorb moisture before testing. Because of the extremely low permeability of shale, saturation degree is not controlled in these experiments and soaking may only make the shale more wet than the room-dry sample.

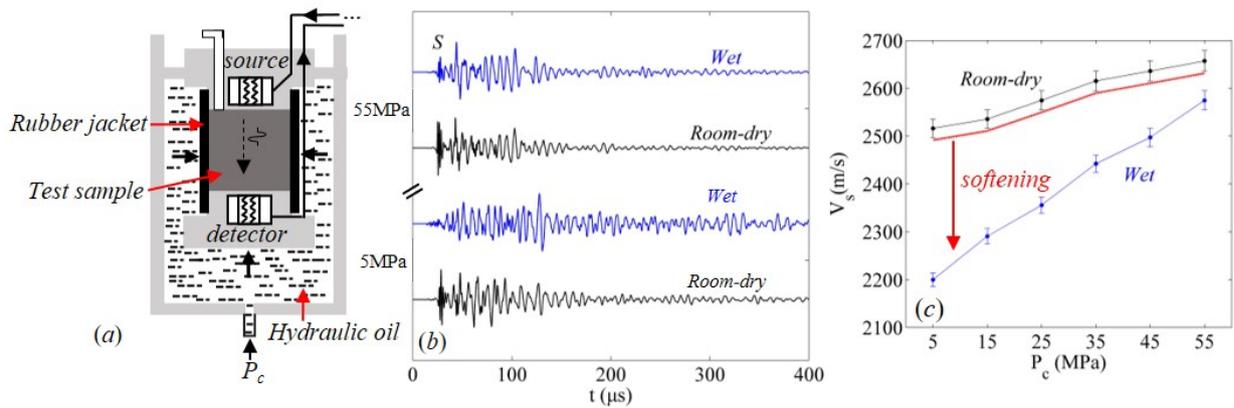

*Figure 2. Ultrasonic experiments in the shale sample under hydrostatic loading. (a) Setup for ultrasonic experiments combined with hydrostatic loading. (b) S-wave signals of room-dry (in black) and wet shales (in blue) in the presence of micro-cracks (cf Fig. 1a) with the confining pressure $P_c$ = 5 MPa and 55 MPa, respectively. (c) S-wave velocities of room-dry and wet shales determined from (b) as a function of $P_c$. For comparison, we show the shear velocity change (red dash curve) predicted by Gassmann [Bourbié et al, 1987]] only due to density change Δρ (~ 2%), which is much smaller than the measured change in $ΔV_s$ (~ 10%).*

The main results of the hydrostatic loading tests are shown in Fig. 2b-2c. In Fig. 2*b*, the scattered waves (coda) centered at $f_c$ = 0.5 MHz are stronger in the *wet* shale at low confining pressure ($P_c$ = 5 MPa) and weaker in the *room-dry* shale at high pressure ($P_c$ = 55 MPa). The difference in ultrasonic coda waves for the *room-dry* sample recorded respectively at 5 MPa and 55 MPa is not significant, while it becomes visible for the wet shale sample. This observation suggests the internal structure change by wetting, which induces additional heterogeneities like micro-cracks. Increasing the confining pressure may close such micro-cracks and reduce wave scattering (i.e., coda wave amplitude). Moreover, the rapid increase of the *S*-wave velocity in the wet shale compared to that in the room-dry sample with increasing $P_c$ (Fig. 2c) is consistent with the picture of crack closure. Interestingly, the *S*-wave velocity in the *wet* shale is about $ΔV_s$ ~ 10% lower than that in the *room-dry* shale, a relative difference much larger than the density change Δρ (~ 2%), implying that the solid skeleton of the shale is damaged (due to micro-cracks) and weakened by wetting. A similar phenomenon has been observed in previous works where the water-clay interactions and moisture-induced cracking were observed in shaly sandstones and shales (Han et al., 1986; Wang et al., 2014, 2015; Zhang et al., 2017).

**2.2 Cyclic heating and cooling in room-dry shale**

We have also conducted the cyclic heating and cooling experiments in the room-dry shale to gradually fracture the sample and monitor its structure change by the ultrasound transmission. In each cycle of heating and cooling, the sample is heated at 105 °C for 24 hours, then quenched in water (wrapped by the waterproof film) at room temperature for about 10 min before ultrasonic measurements. The acoustic setup is shown in Fig. 3a. A 5-cycle sine burst signal centered at 0.5, 1.0, 1.5, and 2.0 MHz, respectively, was amplified (by 50 dB) and sent to a large source transducer (12mm in diameter), longitudinal (*P*) or shear (*S*). The transmitted ultrasound was detected at another end of the sample by the same large transducer for wave velocity measurements, or by a small transducer (2 mm in diameter) to record scattered ultrasound (coda waves).

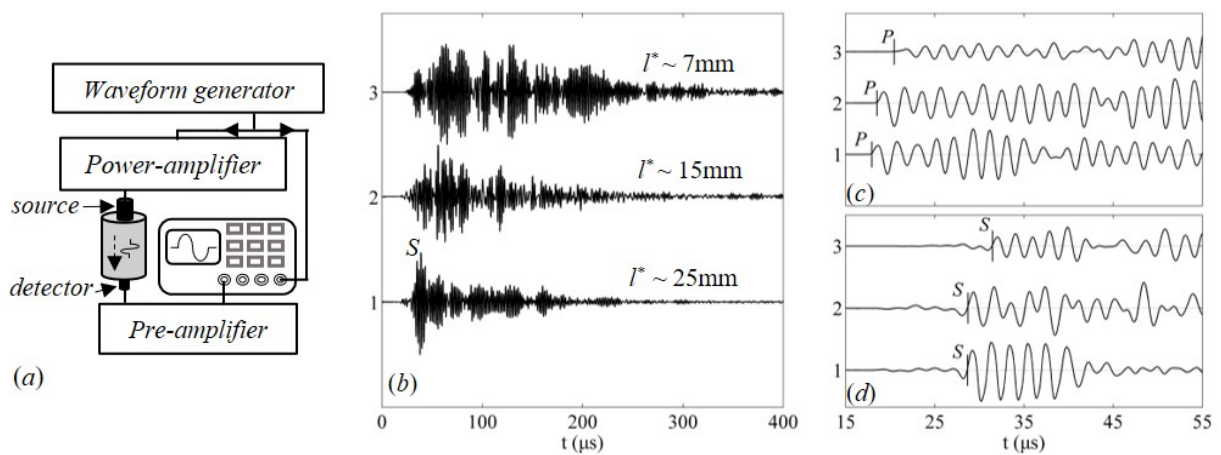

*Figure 3. Ultrasonic experiments in the fractured shale sample at ambient temperature and pressure. (a) Setup of the ultrasonic experiment for monitoring the fractured sample. (b) Typical coda wave signals obtained with a large shear-wave source and a small detector. S denotes the ballistic S-wave and the numbered signal traces "1-3" indicates the sequence of experiments with"1" referring to the initial shale, "2" to the intermediate and "3" to the final fractured shale, respectively. (c) Typical signals obtained from the same samples with a large detector (longitudinal transducer), where P denotes the ballistic P-wave. (d) Typical signals obtained from the same samples but with a large shear transducers where S denotes the ballistic S-wave.*

Fig. 3b-3d depict the main experimental results. Three typical signals centered at 0.5 MHz (Fig. 3b) were excited by a large shear transducer detected by the small transducer, showing the evolution of ultrasonic coda waves transmitted through the different fractured stage of the shale. The coda waves become stronger and stronger from the signal trace "1" to the trace "3" when the fracture develops. Also, the *P*-wave velocity decreases accordingly from 3700 m/s to 3250 m/s (Fig. 3c) and the *S*-wave velocity from 2300 m/s to 2100 m/s (Fig. 3d). This corresponds a reduction of *P*-wave and *S*-wave velocities by 12% and 9%, respectively. In fact, after 8 months of repeated heating and cooling, a number of macro-fractures appeared on the surface of the sample (Fig. 1b1). The comparison of the SEM

photographs of the fractured sample (Fig. 1b2) with those of the initial intact sample (Fig. 1a2) indicates two main types of cracks, one originates from the break of brittle quartz grains (type I, intragranular) and the other arises from mineral boundaries (type II, intergranular). Both types of microcracks have been observed previously in thermal-stressed granite and sandstone (e.g., Lin 2002; Wang et al., 2013). These microcracks grow and interconnect in specific directions (e.g., along the bedding plane), forming millimeter-scale macro-fractures.

Although the same sample was used for both experiments, we must realize that the shales before and after the cyclic heating and cooling are definitely different. For the hydrostatic loading experiments, the sample was relatively intact and ultrasound scattering was likely due to moisture-induced microcracks and softened clays. For the fractured shale by the cyclic heating and cooling, scattering is likely controlled by developed macro-fractures.

## 3. FD simulations and inverse process

To investigate the fracture process, we scanned the shale sample after long cycles of heating and cooling with high-resolution X-ray computed tomography (CT). Fig. 4a shows the raw grayscale CT image of the macro-fractured top surface (Fig. 1b1) with a spatial resolution of 27.5 μm. According to the relative occurrence of gray values, we manually binarized the image with a threshold as shown in Fig. 4b, obtaining an image composed of solids and macro-fractures shown in Fig. 4c. The comparison between Fig. 4a and Fig. 4c confirms the identification of most macro-fractures by this image analysis even though some tiny blurred fractures or cracks in Fig. 4a are either discontinuous or missing. More accurate segmentation of rock CT images requires further efforts to combine manual segmentation processing and advanced segmentation algorithms (Iassonov et al., 2009; Andrä et al., 2013). Note that microcracks observed by the SEM in Fig. 1b2 are not detectable by such CT images due to the resolution limit.

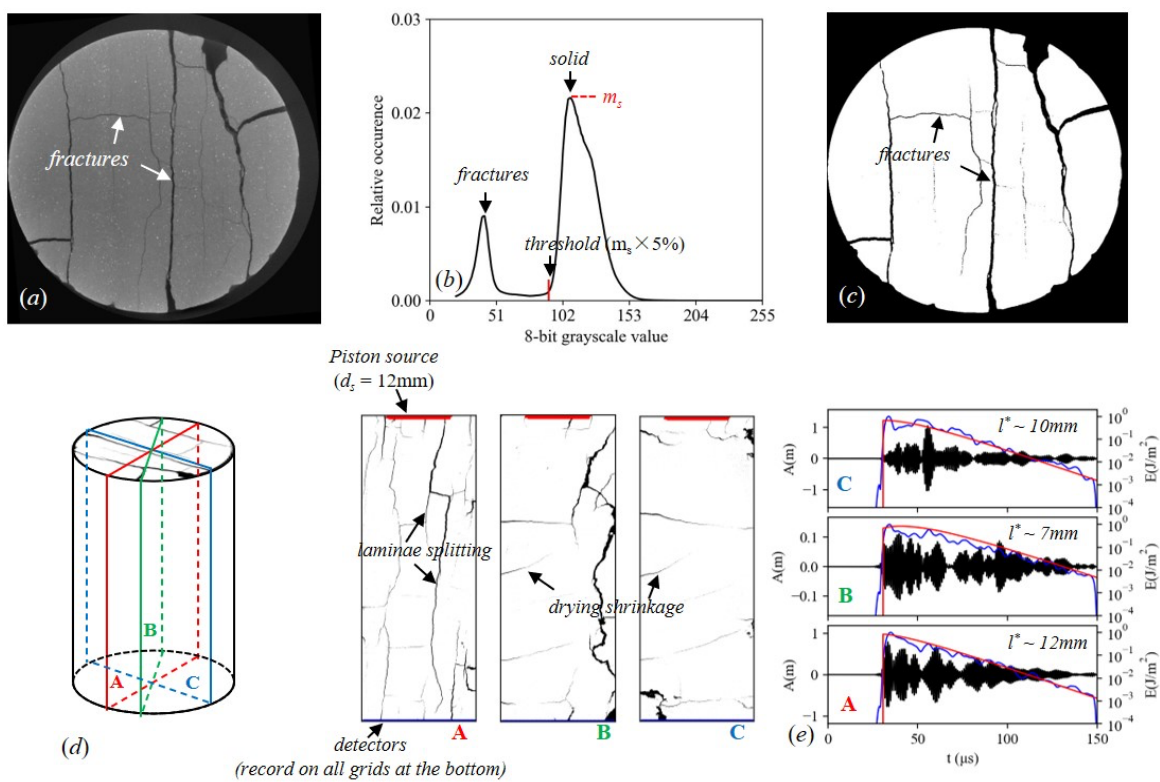

*Figure 4. FD simulations of SH wave scattering in fractured shales. (a) the raw grayscale CT image of the top surface of the final shale sample with macro-fractures. (b) Thresholding the grayscale CT image according to the relative occurrence of gray values. (c) The segmented CT image in terms of solid (white area) and fractures (black area). (d) Three axial sections (A, B, C) of the fractured shale. (e) Typical simulated scattered waves, i.e., coda (black) through sections A, B and C with averaged temporal energy profiles (blue). Both absorption $Q_i^{-1}$ and transport mean-free path l\* may be inferred by fitting the diffusion solution (red), see the text.*

Fig. 4d shows internal fractures through three axial sections, perpendicular to (A), oblique to (B), and parallel to the bedding (C), respectively. In axial section A, we see a few macro-fractures, developed parallel to the bedding plane (local splitting of laminae), bridged by fractures perpendicular to it, running through almost the entire sample. This fracture development pattern is controlled by the material texture and the orientation of the indentation to bedding, as shown by mechanic-cracking experiments in shales combined with synchrotron imaging (Ma et al., 2021). In addition, there are some fractures extending from surface to interior (in axial sections B and C), perpendicular to beddings, showing typical fractures induced by drying shrinkage (Yao et al., 2014), likely due to our cyclic heating and cooling.

### 3.1 Simulation of SH-wave scattering by macro-fractures

Since the length of these fractures is comparable to the S-wave wavelength (~ 2mm), we simulated SH-wave propagation on these axial sections using a finite difference (FD) method, applying a fourth-order velocity-stress scheme with a staggered grid (Levander, 1988; Suzuki et al., 2006). The solid was treated as a homogeneous medium with a density of $\rho_s$ = 2550 kg/m$^3$ and a shear wave velocity $V_s$ = 2000 m/s, the grid spacing $\Delta x$ = 27.5 μm, and the time interval $\Delta t$ = 5 × 10$^{-9}$ s. By setting the shear stress to zero, the four boundaries of a rectangular image were treated as free surfaces. By setting the shear wave velocity to zero for those grid points in cracks, i.e., applying the *vacuum formulation* (Graves, 1996), we treated the cracks as cavities with stress-free boundaries. The density was not set to zero to avoid instability. We have validated the algorithm by applying it to a classical SH-wave scattering by a cylindrical cavity (Mow and Pao, 1971). For ultrasound with a central frequency of $f_c$ = 1 MHz, the grid size is about 1/70 of the wavelength (less than 1/60), so, the staircase scattering is negligible (Bleibinhaus and Rondenay, 2009). A piston-like line source (12 mm in length) is set on one side of the sample, radiating a 5-cycle sine burst tone centered at 1MHz, and signals are recorded on the other side (Fig. 4d). A spatially uniform absorption, characterized by a quality factor $Q_i$, was added at each time step by multiplying the updated velocity and stress fields with a decay factor $F = \exp(-\pi f_c \Delta t / Q_i)$ (Graves, 1996). To gain computing time, we here set $Q_i$ = 100 and recorded the signals with a maximum duration of 150 μs.

### 3.2 Simulation of SH-wave scattering by micro-cracks

In the initial shale sample, there is no macro-fractures (Fig.5a). The heterogeneity in the wet shale is attributed to numerous microcracks created by the swelling of soft clay mostly at the boundary between stiff quartz grains and soft clay, with crack openings around 1 μm and variable lengths up to 50 μm (Wang et al., 2014). To investigate ultrasound scattering by such microcracks, we generate numerically a medium containing a large number of random microcracks and simulate the SH-wave scattering inside it, as did in previous works (e.g., Saenger and Shapiro, 2002; Krüger et al., 2005; Suzuki et al., 2006). In Fig. 5b, there are 84500 micro-cracks in the rectangular medium with a length of 65 mm and a width of 25 mm (crack length is 100 μm and crack width is 10 μm), and the total crack porosity is 5%. For the simulation, $\Delta x$ = 5 μm, $\Delta t$ = 1.25 × 10$^{-9}$ s, $f_c$ = 1 MHz, $Q_i$ = 100, $\rho_s$ = 2550 kg/m$^3$, $V_s$ = 2000 m/s, the algorithm, the boundary condition, and source-receiver settings are the same as described above.

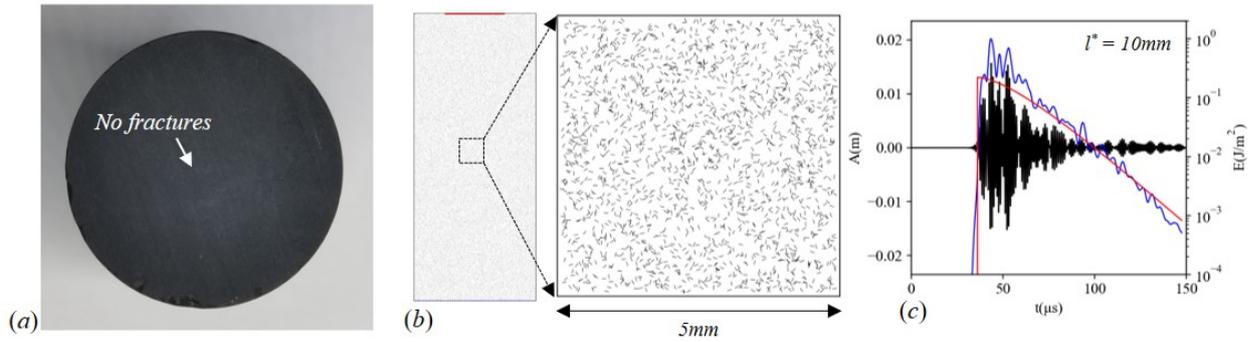

*Figure 5. FD simulations of SH wave scattering by micro-cracks. (a) A photograph of the top surface of the initial intact shale sample (cf Fig. 1a). (b) A two-dimensional medium consisting of randomly distributed and oriented microcracks (100 μm in length and 10 μm in width) generated for simulations in the sample (a). (c) Simulated scattering signal (black) and averaged temporal energy profile (blue) fitted by the diffusion solution (red).*

### 3.3 Inference of wave transport parameters

For a heterogeneous medium with microcracks, we may consider it as a first approximation as a weak-fluctuating elastic random medium, with a structure characterized by a certain correlation function (Klimeš, 2002). Wave scattering in such a medium can be treated with statistical methods, such as radiative transport or diffusion theory (Margerin, 2005; Sato et al., 2012). For a medium with macro-fractures (even not completely random), the wave problem is reminiscent of reverberant sound fields in room acoustics (Kuttruff, 2016). The irregular-shaped zones of our shale separated by macro-fractures may be regarded as chaotically reverberating cavities connected by opened acoustical paths. A diffusion model is an effective approximation for predicting sound energy distribution and its decay in such scattering systems (Valeau et al., 2006; Billon et al., 2006). For simplicity, here we seek to infer the ultrasound transport mean-free path and absorption in the shale via the solution of the diffusion equation in a cylinder geometry with a plane-wave source and perfect reflection boundary condition (Jia, 2004). To retrieve scattering and absorption parameters, the temporal energy density profile is calculated by $E(t) = [f(t)]^2 + \{H[f(t)]\}^2$, where $f(t)$ is the recorded signal and $H[\cdot]$ is the Hilbert transform, and then compared to the diffusion solution. Note, for experiments with only one recording, we assume ergodicity (Wegler and Lühr, 2001) and simply smooth $E(t)$. For simulations with many recordings, we average the energy profile over detectors and configurations.

In Fig. 4e, we depict the simulated waveforms detected at the center of the cylinder on the bottom surface for the frequency range 0.8 MHz < $f$ < 1.2 MHz through different axial sections. These three waveforms are similar to those observed in the mostly fractured sample after the heating-cooling cycle, *i.e.*, the signal trace "3" in Fig. 3b where the waveform is dominated by the multiply scattered waves. Moreover, the solution of the above diffusion model appears to fit reasonably well with the averaged temporal energy density profiles, from which we may deduce a transport mean free path $l^*$ = 10 ± 3 mm, close to the inversed value from the experimental data (Fig. 3b). Such ultrasound scattering is qualitatively consistent with sound reverberation in room acoustics where the transport mean free path $l^*$ is roughly in the order of the chaotic cavity (room) size. There is nevertheless a notable difference in the damping of the coda where $Q_i$ (= 100) is about half of the value inferred from the experimental data (see Fig. 6a).

Fig. 5c shows a similar simulated strong scattering of SH-wave by microcracks. By fitting the averaged energy profile with the diffusion model, we infer a transport mean-free path $l^*$ ~ 10 mm (Fig. 5b). This simulation which agrees qualitatively with our experimental observation supports the scenario of strong ultrasound scattering in the initial (wet) shale sample by a large amount of moisture-induced microcracks. Generally speaking, we need to consider both P- and S-waves scattering as occurred experimentally in real shale samples. Because of the mode conversion, multiple scattered elastic waves shall be dominated by S-wave at the long-time range (Weaver, 1990; Aki,1992; Hennino et al., 2001). Therefore, simplified 2D simulations performed here by using the *scalar-like* SH wave provide an efficient tool to investigate the ultrasonic scattering in fractured shales.

## 4 Results and Discussion

### 4.1 Ultrasound scattering by micro-cracks and macro-fractures

The above numerical simulations and experiments have shown that a fractured medium filled with many micro-cracks or with much less macro-fractures may result in the similar strong wave scattering with a comparable transport mean-free path $l^*$. To further investigate the issue, we compare in Fig. 6a the scattering attenuation $Q_s^{-1}$ ~ $V_s/(\omega l_s)$ (with $l_s$ the scattering mean-free path and $V_s$ the shear wave velocity) predicted in exponential random media (Sato et al, 2012) and that inferred from coda waves in the initial shale (Fig. 1a1) and in the final fractured sample (Fig. 1b1), respectively, after the heating-cooling cycles. The isotropic scattering approximation is assumed here with $l_s$ ~ $l^*$ and the frequency is rescaled by $ka$ with $k = \omega/V_s$ ($V_s$ = 2000 m/s) and $a$ the correlation length (an adjustable parameter) previously introduced in exponential random media where the scattering attenuation $Q_s^{-1}$ is given by,

$$Q_s^{-1} = \frac{2\varepsilon^2(ka)^3(4-v_c^2)}{(1+v_c^2(ka)^2)(1+4(ka)^2)} \qquad (1)$$

with $\varepsilon$ the wave velocity fluctuation, and $v_c$ (= ¼, also a fitted value) related to the cutoff angle (Kawahara, 2002). In Fig. 6a1, the inversed $Q_s^{-1}$ of the initial *room-dry* shale increases with frequency from 0.5 MHz (signal trace "1" in Fig. 3b, $ka$ = 0.4, $Q_s$ = 250 ± 50) to 2 MHz ($ka$ = 1.7, $Q_s$ = 40 ± 5). By a best fit of the inversed $Q_s^{-1}$, we obtain $\varepsilon$ = 10% and $a$ = 0.3 mm, suggesting that all inversed $Q_s^{-1}$ lie in the Rayleigh scattering regime with the effective scatterer size smaller than the wavelength $\lambda$ ~ 1 mm. As a result, the exponential random medium in Fig. 6b1 could mimic the structure of the initial shale where the wave velocity fluctuation originates likely from microcracks at boundaries between clay and quartz grains (Fig. 6c1). For the *wet* shale, scattering is even stronger ($Q_s$ = 20 ~ 40 for *f* = 0.5 MHz) due to the clay softening by wetting, inducing additional micro-cracks (Wang et al., 2014, 2015).

In opposite, the $Q_s^{-1}$ inferred from the final shale with macro-fractures (Fig. 6a*2*) decreases from 0.5 MHz (signal trace "3" in Fig. 3b, *ka* = 5.5, $Q_s$ = 14 ± 2) to 2 MHz (*ka* = 22, $Q_s$ = 35 ± 5). By fitting, we get $\varepsilon$ = 14%, *a* = 3.5 mm; all inverted $Q_s^{-1}$ are situated on the other side of the scattering curve, indicating a forward scattering regime in which the correlation length (effective scatter size) is larger than the wavelength. From the wave scattering viewpoint, the macro-fractured shale could then be approximated by an exponential random medium with the wave velocity fluctuation on the millimeter scale shown in Fig. 6b2. This structure is consistent with CT images of the large fractures in Fig. 4 due to the growth of micro-cracks sketched in Fig. 6c2.

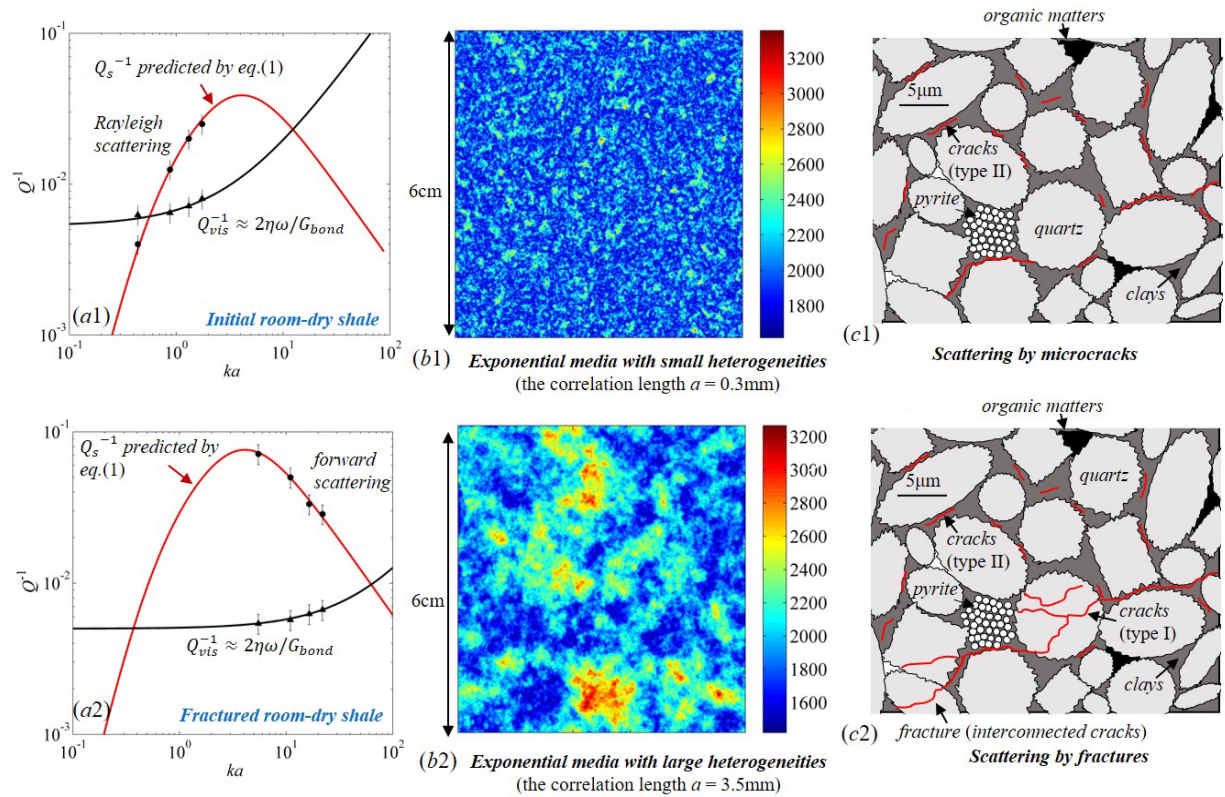

*Figure 6. Comparison of scattering attenuation between fractured shales and exponential random media. (a1) The scattering attenuation (circle) and intrinsic attenuation (triangle) measured for the frequency ranged from 0.5 MHz to 2.0 MHz in the initial shale (room-dry) with k the wave number and the correlation length (see text). The red curve corresponds to the prediction in an exponential random medium. (b1) An exponential random medium (model) generated with correlation length a = 0.3 mm where red areas refer to large velocities while blue areas refer to small velocities. (c1) The sketch of the initial shale with microcracks. (a2) The scattering attenuation (circle) and intrinsic attenuation (triangle) for the frequency ranged from 0.5 MHz to 2.0 MHz in the final shale (room-dry). (b2) An exponential random media with correlation length a = 3.5 mm. (c2) The sketch of the final shale with macro-fractures interconnecting intergranular and intragranular microcracks.*

As a conclusion, we believe that ultrasound scattering in the initial shale and the final fractured shale is different: the former is controlled by dense microcracks while the latter is mainly due to dilute large fractures. Within the framework of the wave scattering in exponential random media, our experimental observations suggest an increase in the correlation length *a*, indicating the nucleation process of micro-cracks up to failure via macro-fractures by cyclic heating-cooling. The increase in the fluctuation magnitude *ε* could also be an indication of the progressively fractured (damaged) medium.

### 4.2 Mechanisms of coda wave absorption

Figs. 6a1 and 6a2 show that the intrinsic attenuation (absorption) $Q_i^{-1}$ is generally smaller than the scattering attenuation in room-dry shales for the ultrasonic frequency range considered here. Similar values of $Q_i^{-1}$ are also inferred in the initial wet shale from the hydrostatic loading experiments (see below Fig. 8a) using the large transducer detection. To examine the possible effect of the detector size on scattering measurements, we compare in Fig. 7a ultrasound transmissions in the initial room-dry shale detected by a large transducer (blue trace) and a small one (black trace), respectively, together with their averaged energy density profiles shown in Fig. 7b. The large size detector enhances clearly the ballistic wave but reduces the multiply scattered waves by the incoherent interferences (Page et al, 1995; Jia et al, 1999). Nevertheless, such a spatial average doesn't affect the absorption measurement based on the exponential decay of coda waves at the long-time range (Fig. 7b) provided the S/N ratio of scattered waves remain high enough. Instead, it may induce an apparent increase of the scattering mean-free path $l_s$ (~ $l^*$) or the $Q_s$ (from 50 to 100) which is inferred mainly from the rising edge of the energy density profile at the short time via the radiative transfer equation or the diffusion model (e.g., Zhou et al 2021).

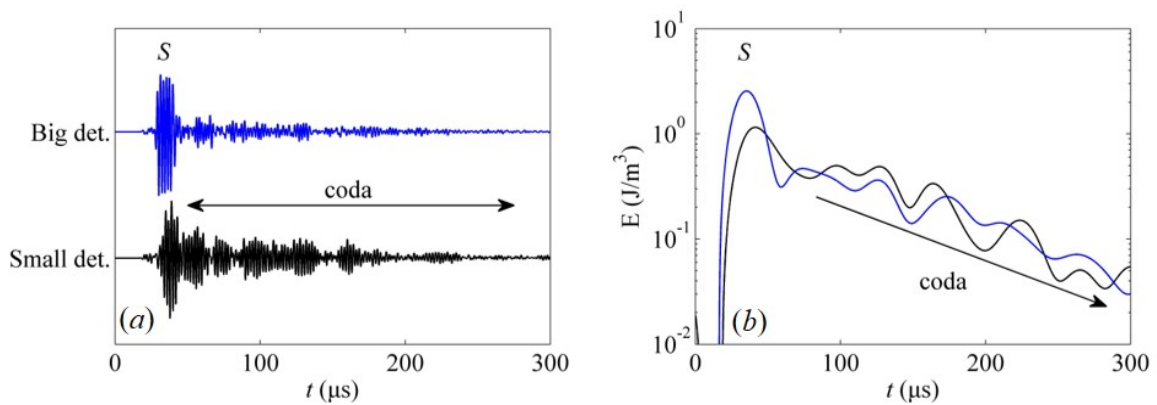

*Figure 7. (a) Ultrasonic coda waves recorded by a big detector(blue) and a smaller detector(black). (b) Temporal energy density profiles from a large detector (blue) and a smal detector (black).*

In addition to the linear increase of $Q_i^{-1}$ with the frequency observed in the room-dry shale (Figs. 6a1 and 6a2) as expected in viscoelastic materials (the linear dependence may not obvious in the log-log plot), we now investigate the intrinsic dissipation $Q_i^{-1}$ and the scattering attenuation $Q_s^{-1}$ as a function of the confining pressure $P_c$ (Fig. 8a) with a large detector discussed above. If the decrease of $Q_s^{-1}$ measured with increasing $P_c$ can be expected (despite of the aveargage effect on the quantitative value of $Q_s$ mentioned above) due to the closure of micro-cracks induced by the swelling of wetted clays (Figs. 2b and 2c), the increase of $Q_i^{-1}$ is somehow surprising. Indeed, this contradicts the expectation based on the well-known wave-induced fluid flow models which predict a decrease of $Q_i^{-1}$ with increasing effective pressure in porous rock materials (e.g., Jones, 1986; Winkler and Murphy, 1995).

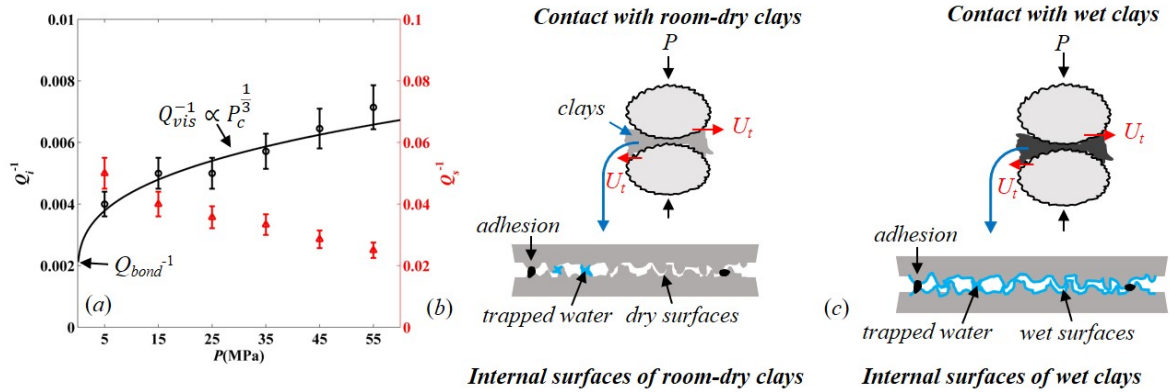

Figure 8. Pressure-dependent intrinsic attenuation $Q_i^{-1}$ (black circles) and scattering attenuation $Q_s^{-1}$ (red triangles) in the wet shale with micro-cracks (black points). Inferred $Q_i^{-1}$ are compared with a heuristic model based on the frictional dissipation between grain (quartz) contact, viscous dissipation of trapped liquid films and/or clay around rough grain-grain contacts, as well as cementation or bonding of two quartz particles through wet clays (sketches).

Here we propose a scenario to interpret this main finding, along the line of ultrasound scattering in highly confined dense granular media (Jia et al, 1999; Jia, 2004). Similar to the drastic acoustic attenuation associated with small amounts of volatiles tightly bound to the host particle surface (Tittmann et al., 1980), Brunet et al. (2008) have found that elastic wave dissipation in weakly wet granular media is dominated by a viscous loss in the liquid film trapped by asperities at the rough grain surface. It leads to a high shear rate induced by the wave, $d\gamma/dt = \omega\gamma \sim (2\pi f)(U_t/\delta)$ with $f$ the ultrasound frequency, $U_t$ (~ 1 nm) the shear wave amplitude and $\delta$ (~ 10-100 nm) the thickness of adsorbed films. Fig. 8b depicts a sketch of two quartz particles in contact surrounded by intergranular clay minerals in a room-dry shale, more or less wetted by adsorbed water. When a shear wave travels through the solid frame from grain to grain, the ultrasound absorption $Q_i^{-1}$ could be similarly considered as the addition of a viscous dissipation $Q_{vis}^{-1}$ due to adsorbed water and an interfacial solid friction $Q_{fric}^{-1}$ within a mean-field description,

$$Q_i^{-1}(P_c,\omega) = Q_{vis}^{-1}(P_c,\omega) + Q_{fric}^{-1}(P_c) \qquad (2)$$

where $Q_{vis}^{-1}(P_c,\omega)$ is a linear viscous loss dependent of frequency $f$ and confining pressure $P_c$ due to the increased contact area (or closure of cracks), and $Q_{fric}^{-1}(P_c)$ is a nonlinear frictional loss dependent of $U_t$ and $P_c$ but independent of $f$. Accordingly, we may estimate $Q_{fric}^{-1}$ from the measurement of $Q_i^{-1}$ in room-dry shales with the extrapolation at zero-frequency (Figs. 6a1 and 6a2). The deduced value is about $Q_{fric}^{-1} \sim 5\times10^{-3}$, comparable to those in unconsolidated granular materials (Brunet et al, 2008).

In a wet shale (Fig. 8c), we speculate that the internal surface of the clay minerals is mostly covered by a water film where the viscous dissipation becomes dominant $Q_{vis}^{-1} \gg Q_{fric}^{-1}$ (Brunet et al, 2008). Note however that unlike wet granular packings, the presence of clay minerals between the quartz particles induces strong adhesion and can create the bonds between solid particles (Digby, 1981; Winkler, 1983; Dvorkin et al, 1994, 1999) as sketched in Fig. 7b and Fig. 7c. The dissipation associated with these viscoelastic bonds should be added to the viscous loss caused by the trapped liquid films in clays or/and confined between grains,

$$Q_{vis}^{-1}(P_c, \omega) = Q_{liq}^{-1}(P_c, \omega) + Q_{bond}^{-1}(\omega) \quad (3)$$

More specifically, we may calculate the dissipation by $Q^{-1}$ = (1/2π) $E_d$/$E_s$ with $E_d$ the energy dissipated per cycle and $E_s$ the energy stored per cycle, caused by the trapped liquid and the adhesive bond, respectively. Eq. (4a) depicts the calculated shear wave dissipation $Q_{liq}^{-1}$ due to the trapped liquid film of thickness $\delta_0$, extended over a Hertz contact area $\Sigma_0$ between two spherical particles loaded by the confining pressure $P_c$ (Brunet et al, 2008). For the adhesive bond, we may assume a similar geometry of disc with thickness $\delta_0$ and bonded area $\Sigma_0$. More specifically, we have $E_s$ = (1/2) $k_{bond} U_t^2$ where $U_t$ is the amplitude of shear wave displacement and $k_{bond}$ = $G_{bond}\Sigma_0/\delta_0$ is the shear stiffness of the bonding with $G_{bond}$ the shear modulus of clays. The viscous dissipation in this thin bond is evaluated by $E_d$ = ($\sigma_{bond} \Sigma_0$)$v_t$ (2π/$\omega$) where $\sigma_{bond}$ = $\eta_{clay}(d\gamma/dt)$ is the shear stress with $\eta_{clay}$ the dynamic viscosity, $d\gamma/dt = \omega\gamma \sim \omega(U_t/\delta)$ is the shear rate and $v_t = \omega U_t$ is the shear vibration velocity. We obtain finally $E_d \sim 2\pi\eta \Sigma_0 \omega U_t^2/\delta_0$ and a viscoelastic loss $Q_{bond}^{-1}$ given in Eq. (4b),

$$Q_{liq}^{-1} \approx \frac{(2-\nu)\pi R \eta_{liq}}{2G\delta_0}\left(\frac{3\pi}{4E^*}\right)^{\frac{1}{3}} \omega P_c^{\frac{1}{3}} \quad (4a)$$

$$Q_{bond}^{-1} \approx \frac{2\eta_{clay}}{G_{bond}}\omega \quad (4b)$$

with $E^*$= $E/(2-2\nu^2)$ with $E$, $G$ and $\nu$ the Young, shear moduli and Poisson ratio of solid grains, $R$ the mean grain radius, $\eta_{liq}$ and $\eta_{clay}$ the viscosities of adsorbed liquid and clay.

In Fig. 8a, we compare the prediction by the above model (Eqs. 3 and 4) with the inversed absorption $Q_i^{-1}$ as a function of the confining pressure. The agreement appears fairly well with the reasonable fit parameters: $R \sim 10$ μm, $\delta_0 \sim 5$ nm, $E$ = 100 GPa, $G$ = 45 GPa, $E^*$= 50 GPa, $\nu$ = 0.067 and $\eta_{liq} \sim 0.1$ Pa·s for a liquid in highly confined spaces (van den Wildenberg et al., 2019). Therefore, the *astonishing* increase of absorption with increasing confining pressure $P_c$ could be explained by an increase of the contact area between grains covered by adsorbed water predicted by the Hertz-Mindlin contact theory (Eq. 4a); they are highly sheared by the wave when propagating through the contact network (solid frame). Finally, the finite damping observed at vanishing pressure $P_c \rightarrow 0$ may be ascribed to the viscoelastic dissipation of the internal cohesive bonds (Eq. 4b) at or near grain contacts.

### 4.3. Weakening of shales by wetting

As mentioned above, the phyllosilicate (sheet silicate) nature of clays is responsible for its strong cohesion via the electrostatic interaction. Wetting/humidity or/and thermally induced cracks may however affect the cohesion of clays and weaken its elastic modulus like $G_{bond}$. Detailed investigation of the cohesion in shales is beyond the scope of this work and needs further study in the future. Here we seek to compare the bond modulus in room-dry and wet shale samples inferred with Eq. (4b), indicating that the stronger the bond, the lower the absorption is.

For the *room-dry* shale, we may infer $G_{bond} \approx 1.0$ GPa and $G_{bond} \approx 1.5$ GPa from the slope of the frequency-dependent $Q_i^{-1}$ (Eq. 4b) in Fig. 6*a*1 and Fig. 6*a*2, respectively, with $\eta_{clay} \sim 0.1$

Pa·s. All these values are consistent with the viscoelastic property of clay minerals (Bourbié, 1986; Mavko et al., 2009). On the other hand, we may also deduce $G_{bond}$ ~ 0.4 GPa for the initial *wet* shale in Fig. 8a through the inferred dissipation $Q_i^{-1}$ using Eq. (4b) at $P_c \rightarrow 0$ with the same viscosity $\eta_{clay}$. The comparison of these shear moduli $G_{bond}$ inferred from the dissipation measurement between room-dry and wet shale illustrates a weakening of the shear modulus (from 1.5 GPa to 1 GPa then to 0.4 GPa) when increasing the water content. This observation is well consistent with the softening of shear wave velocity measured in the wet shale sample (Fig. 2c), supporting thus the validity of the absorption determination by our coda wave measurements in fractured shales.

## 5. Conclusion

We have performed ultrasonic experiments and simulations to monitor the structural change and damage in shale samples by wetting and by thermal loading, respectively. Ultrasonic waves are strongly scattered in these shales by micro-cracks due to moisture-induced swelling of clays or by macro-fractures caused by cyclic heating and cooling. The increase of ultrasound scattering with the material damage corroborates with the observation of P- and S-wave velocity decrease (TenCate et al, 2002; Fortin et al, 2007; Simpson et al, 2022). We find that increasing the confining pressure enhances wave velocities, likely by crack closures, but also the internal dissipation. This main finding, opposite to the prediction by mechanisms based on the wave-induced fluid flow, stems from the viscous dissipation of small amount of liquids (water) adsorbed in clays or trapped by rough grain surfaces. This observation is consistent with the modified Hertz-Mindlin contact model, including the adhesive bonds.

Thanks to the X-ray CT imaging of the fractured shale sample by thermal loading, we reconstructed numerically a heterogeneous medium with macro-fractures and simulated the shear ultrasound scattering through the medium. FD simulations of the shear ultrasound scattering were also performed in a model system filled with many micro-cracks. Multiply scattered waves simulated in both heterogenous systems agree fairly well with the diffusion model, which allows us to invert the intrinsic and scattering attenuations (or mean-free path) in both the fractured shales. Moreover, we have compared these scattering mean-free paths with those predicted in the *exponential random medium*. The reasonably good agreement enables to extract a characteristic correlation length for our heterogeneous system, which increases consistently from the initial intact shale with micro-cracks to the final fractured sample with macro-fractures. This interesting and useful result deserves further investigation towards the definition of a possible indicator for material damage which allows the better understanding of fracture nucleation.

Note that possible anisotropic effects caused by oriented bedding on wave scattering and dissipation should be considered in the future. Dominant ultrasound scattering attenuation revealed in this work for heterogeneous rocks like shales ($Q_s^{-1} > Q_i^{-1}$) suggests that the absorption measurement based on coherent ultrasound propagation needs to be carefully analyzed by taking into account the scattering attenuation. We believe this study helps to improve acoustic techniques for gas and oil exploration in shales and other fractured rocks.

**Acknowledgments**

The authors are grateful to Prof. Ji-Xin Deng at the Chengdu University of Technology, China for providing shale samples. We thank Prof. Haruo Sato for the fruitful discussion about the scattering mechanism. We acknowledge the Institute of Geochemistry in the Chinese Academy of Sciences for the visiting scholarship of H.Z. and the Institut Langevin of ESPCI Paris - PSL for the research assistance. This work has received support from PSL University and under the program "Investissements d'Avenir" launched by the French Government.